\documentclass[aps,prl,twocolumn,superscriptaddress]{revtex4-1}

\usepackage{graphicx}% Include figure files

\begin{document}

%\preprint{}

\title{Pressure-Induced Valence Crossover and Novel Metamagnetic Behavior near the Antiferromagnetic Quantum Phase Transition of YbNi$_{3}$Ga$_{9}$}

% address or url should go in the {}'s for \email and \homepage.% Please use the appropriate macro foreach each type of information

%\author{}
%\thanks{}
%\affiliation{}

\author{K.~Matsubayashi}
\email{kazuyuki@issp.u-tokyo.ac.jp}
\affiliation{Institute for Solid State Physics, The University
of Tokyo, Kashiwanoha, Kashiwa, Chiba 277-8581, Japan}
\author{T.~Hirayama}
\affiliation{Institute for Solid State Physics, The University
of Tokyo, Kashiwanoha, Kashiwa, Chiba 277-8581, Japan}
\author{T.~Yamashita}
\affiliation{Department of Engineering Physics, Electronics and
Mechanics, Graduate School of Engineering, Nagoya Institute of
Technology, Nagoya 466-8555, Japan}
\author{S.~Ohara}
\affiliation{Department of Engineering Physics, Electronics and
Mechanics, Graduate School of Engineering, Nagoya Institute of
Technology, Nagoya 466-8555, Japan}
\author{N.~Kawamura}
\affiliation{Japan Synchrotron Radiation Research Institute
(JASRI/SPring-8), 1-1-1 Kouto, Sayo, Hyogo 679-5198, Japan}
\author{M.~Mizumaki}
\affiliation{Japan Synchrotron Radiation Research Institute
(JASRI/SPring-8), 1-1-1 Kouto, Sayo, Hyogo 679-5198, Japan}
\author{N.~Ishimatsu}
\affiliation{Department of Physical Science, Graduate School of
Science, Hiroshima University, 1-3-1 Kagamiyama,
Higashi-Hiroshima, Hiroshima 739-8526, Japan}
\author{S.~Watanabe}
\affiliation{Quantum Physics Section, Kyushu Institute of Technology, Fukuoka 804-8550, Japan}
\author{K.~Kitagawa}
\affiliation{Graduate School of Integrated Arts and Sciences,
Kochi University, Kochi 780-8520, Japan}
\author{Y.~Uwatoko}
\affiliation{Institute for Solid State Physics, The University
of Tokyo, Kashiwanoha, Kashiwa, Chiba 277-8581, Japan}


\date{\today}

\begin{abstract}
We report electrical resistivity, ac magnetic susceptibility and X-ray absorption spectroscopy measurements of intermediate valence YbNi$_{3}$Ga$_{9}$ under pressure and magnetic field. We have revealed a characteristic pressure-induced Yb valence crossover within the temperature-pressure phase diagram, and a first-order metamagnetic transition is found below $P_{\rm c}$ $\sim$ 9~GPa where the system undergoes a pressure-induced antiferromagnetic transition. As a possible origin of the metamagnetic behavior, a critical valence fluctuation emerging near the critical point of the first-order valence transition is discussed on the basis of the temperature-field-pressure phase diagram.
\end{abstract}

%\keywords{}
%71.27.+a 	Strongly correlated electron systems; heavy fermions
%74.62.Fj 	Effects of pressure

\maketitle

%\section{}
%\subsection{}
%\subsubsection{}

	In heavy fermion compounds, tuning the ground state by pressure and/or magnetic field from a non-magnetic state to a magnetic state or vice versa has attracted attention because the anomalous behavior, such as unconventional superconductivity or non-Fermi liquid (NFL) state, appears in the vicinity of quantum critical point (QCP) where a second order magnetic phase transition is suppressed to $T$ = 0 K \cite{RevModPhys.79.1015,Gegenwart:2008aa}. The conventional spin fluctuation theories reproduce the NFL behavior in many cases \cite{SCR-1,SCR-2,SCR-3}, however, recent studies, especially on Yb-based heavy fermion compounds such as YbRh$_{2}$Si$_{2}$, $\beta$-YbAlB$_{4}$ and Yb$_{15}$Al$_{34}$Au$_{51}$, revealed that these systems exhibit anomalous quantum critical behavior deviating from the conventional QCP scenario and the common low-temperature exponents of the physical properties are observed \cite{PhysRevLett.89.056402,Matsumoto21012011,Deguchi:2012aa}. In particular, an intriguing mystery is an enhanced uniform magnetic susceptibility, giving rise to a large Wilson ratio in spite of the absence of a ferromagnetic phase nearby. To elucidate the nature of the unconventional critical behavior and the underlying physics, a number of theories have been proposed such as the local criticality theory based on the Kondo breakdown QCP, the theory of the tricritical point and the theory of the QCP of valence transition \cite{theory-LQCP-1,theory-LQCP-2,theory-TCP,theory-valence}. Although these theories predict some important aspects across the QCP such as a jump in the Fermi surface volume or a critical valence fluctuation, the nature of the unconventional criticality still remains an open question.
	
	Whereas most of the detailed investigations were carried out by tuning magnetic field, another important clue as to the nature of the quantum criticality has come from high-pressure studies. In ytterbium (Yb) systems, it is well known that the evolution of magnetism from nonmagnetic Yb$^{2+}$ to magnetic Yb$^{3+}$ is achieved through the application of pressure \cite{YbCu2Si2,YbCo2Zn20}. In the vicinity of the magnetic QCP, novel metamagnetic behavior is often observed in the paramagnetic regime \cite{MM_YbCu2Si2,MM_YbCo2Zn20}, possibly related to a change of Fermi surface or valence instability. Therefore, it is natural to raise the question how the metamagnetic transition as well as the Yb valence state evolve approaching to a magnetic QCP. However, experimental complications in combination with high pressure, magnetic field and low temperature make it challenging.
	
	In this Letter, we report a comprehensive study on intermediate valence YbNi$_{3}$Ga$_{9}$ using both hydrostatic pressure and magnetic field as tuning parameter, and present the precise temperature-magnetic field-pressure ($T$-$H$-$P$) phase diagram in the vicinity of the pressure-induced antiferromagnetic (AFM) transition together with the pressure variation of the Yb valence state. YbNi$_{3}$Ga$_{9}$ crystallizes in the ErNi$_{3}$Al$_{9}$-type layered structure \cite{Gladyshevskii}. In this structure, Yb-ions are in Yb$_{2}$Ga$_{3}$-layer separated by seven nonmagnetic triangular-layers of Ga or Ni-ions and form a two-dimensional honeycomb-lattice \cite{YbNi3Al9-1,sample_ohara,sample_yamashita}. At ambient pressure, YbNi$_{3}$Ga$_{9}$ shows valence fluctuation behavior with a Kondo temperature $T_{\rm K}$ of 570~K. In contrast to the paramagnetic magnetic ground state in YbNi$_{3}$Ga$_{9}$, the isostructural YbNi$_{3}$Al$_{9}$ exhibits a helical magnetic order at $T_{\rm M}$~$\sim$~3.4~K with the propagation vector ${\it \bf k}$ = (0, 0, 0.8) \cite{sample_yamashita,Chiral,Metamag_YbNi3Al9,neutron}. Recent X-ray photoemission spectroscopy reported that the Yb valence states of YbNi$_{3}$Ga$_{9}$ and YbNi$_{3}$Al$_{9}$ at low temperature were estimated to be 2.43 and 2.97, respectively \cite{HXPES}. Therefore, high-pressure study on YbNi$_{3}$Ga$_{9}$ is expected to cross a magnetic QCP because applying pressure favors the magnetic Yb$^{3+}$ configuration with a smaller volume. Here we demonstrate that the realization of the pressure-induced valence crossover and the metamagnetic behavior in YbNi$_{3}$Ga$_{9}$ near the AFM quantum phase transition, which suggests the relevance of the valence instability for the quantum critical behavior in heavy fermion systems.

\begin{figure}[t]
\begin{center}
\includegraphics[width=0.4\textwidth]{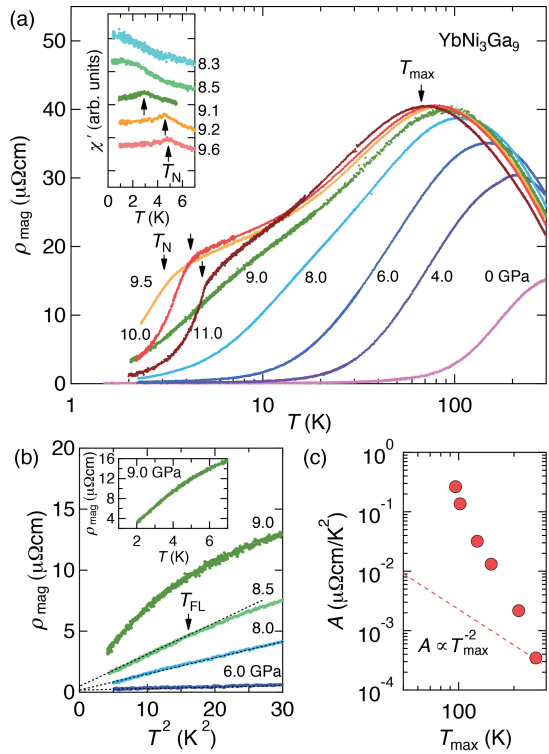}
\end{center}
\caption{(Color online) (a) Temperature dependence of $\rho_{\rm mag}$ of YbNi$_{3}$Ga$_{9}$ under various pressures. The inset shows $T$-dependence of $\chi_{\rm ac}^{'}$ at various pressures.%criterion used to determine TN, The criterion we took to determine the transition temperature is defined a shown in the inset of Fig   
The distinct anomalies as marked by arrows at $T_{\rm N}$ in $\rho_{\rm mag}$ and $\chi_{\rm ac}^{'}$ correspond to the AFM transition. (b) $\rho_{\rm mag}$ as a function of $T^{2}$ at selected pressure. The inset shows the low temperature  $\rho_{\rm mag}$ at 9.0 GPa in a linear scale. (c) The log-log plot of $A$ coefficient and $T_{\rm max}$. The broken line indicates $A$ $\propto$ $T_{\rm max}^{-2}$.}
\label{Fig1}
\end{figure}

\begin{figure}[t]
\begin{center}
\includegraphics[width=0.45\textwidth]{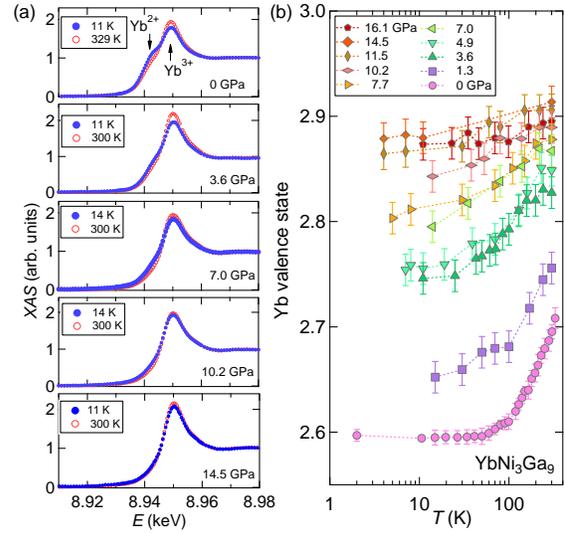}
\end{center}
\caption{(Color online) (a) XAS spectra of YbNi$_{3}$Ga$_{9}$ at selected pressures and temperatures for the Yb~$L_{\rm 3}$-edge. (b) Temperature dependence of the averaged valence for various pressures. }
\label{Fig2}
\end{figure}

	Single crystals of YbNi$_{3}$Ga$_{9}$ were grown by a Ga self-flux method as described previously \cite{sample_yamashita,sample_ohara}. The residual resistivity ratio is 460, reflecting the high quality of the single crystals. Electrical resistivity was measured by a standard four-probe technique with current flow along the $a$-axis using a cubic anvil cell, in which highly hydrostatic pressure is realized owing to the multiple-anvil geometry \cite{exp_cubic}. %The applied pressure inside the cubic anvil cell at room temperature and low temperature is calibrated by the measurement of the resistivity changes of Bi and Te associated with their structural phase transitions and the pressure dependence of the superconducting transition temperature of lead, respectively.
The ac magnetic susceptibility was measured by a conventional mutual-inductance technique at a fixed frequency of 317~Hz with a modulation field of 0.1~mT applied along the $a$-axis. A newly developed opposed-anvil pressure cell with was used for ac susceptibility measurements \cite{exp_KTG}. The applied pressure was calibrated by the pressure dependence of the superconducting transition temperature of lead. X-ray absorption (XAS) measurements at the Yb $L_{\rm 3}$-edge were performed under pressure at the beamline BL39XU of SPring-8, Japan \cite{exp_XAS}. The sample was loaded in a diamond anvil cell (DAC) together with ruby chips, which served as a pressure manometer. Nano-polycristalline diamond (NPD) anvils were used to avoid glitches in XAS spectra \cite{irifune, ishimatsu}. The X-ray wave vector was aligned parallel to the $c$-axis. In the above high-pressure experiments, the pressure-transmitting mediums for the ac magnetic susceptibility and the others (resistivity and XAS) were argon and glycerin, respectively.

	Figure 1 shows the temperature dependence of the magnetic part of the electrical resistivity ($\rho_{\rm mag}$) of YbNi$_{3}$Ga$_{9}$ at selected pressures. Here, $\rho_{\rm mag}$ is obtained by subtracting the resistivity of the isostructural non-magnetic compound LuNi$_{3}$Ga$_{9}$ \cite{sample_yamashita, SA2}. At ambient pressure, $\rho_{\rm mag}$ exhibits a broad peak centered at around room temperature. Application of pressure enhances the magnitude of the $\rho_{\rm mag}$ and the maximum temperature of $T_{\rm max}$ for $\rho_{\rm mag}$ shifts to lower temperature. At pressures above $\sim$9.5~GPa, the pressure dependence of $T_{\rm max}$ tends to be saturated and a new resistive anomaly abruptly appears at $T_{\rm N}$~$\sim$~3~K, which becomes more pronounced and shifts to higher temperature with increasing pressure. As shown in the inset of Fig.~1, ac magnetic susceptibility experiments showed a clear cusp at almost the same temperature as that of the resistive anomaly, indicating the AFM transition at pressures exceeding $P_{\rm c}$ $\sim$ 9.0~GPa.
		
	Pressure variation of the low temperature resistivity toward the $P_{\rm c}$ was analyzed in terms of Fermi-liquid behavior: $\rho$ = $\rho_{\rm 0}$ + $AT^{2}$ at $T$ $\leq$ $T_{\rm FL}$. As shown in Fig.~1(b), the $A$-value is strongly enhanced and $T_{\rm FL}$ shifts to the lower temperatures with increasing pressure, whereas the temperature dependence changes from the $T^{2}$ to a linear behavior in the vicinity of $P_{\rm c}$ (see data at 9.0 GPa in the inset of Fig.~1(b)). Next, we plot $A$ vs $T_{\rm max}$ in a log-log scale in Fig.~1(c). Assuming the $T_{\rm max}$ scales with the Kondo temperature $T_{\rm K}$, $A$ is expected to follow $\propto$ $T_{\rm max}^{-2}$ relationship. However, we observe a clear deviation from the aforementioned relation, suggesting pressure-induced crossover from the weakly correlated to strongly correlated heavy fermion regime due to the valence crossover.
	
	In order to track the pressure variation of Yb valence state, Yb~$L_{\rm 3}$-edge XAS spectra of YbNi$_{3}$Ga$_{9}$ have been measured at various temperatures and pressures as shown in Fig.~2(a). Reflecting the mixed valence character of the Yb ions at ambient pressure, the $L_{\rm 3}$ transitions of Yb$^{2+}$ and Yb$^{3+}$ are observed in the spectra as a weak shoulder at $\sim$8.944 keV and a prominent peak at $\sim$8.950 keV, respectively. The relative intensity of the Yb$^{2+}$ component to that of Yb$^{3+}$ component increases with lowering temperature. With increasing pressure, this temperature variation of the Yb valence state becomes weaker and the spectral weight transfers from Yb$^{2+}$ to Yb$^{3+}$ state. The averaged valence was determined by fitting the XAS spectra to an arctangent step function and a Lorentzian peak for each valence state (see the Supplemental Material \cite{SA1}). As shown in Fig.~2(b), Yb valence monotonically increases with increasing pressure and reaches a value close to 2.9 at around $P_{\rm c}$ where the temperature dependence of Yb valence becomes weak.%At pressures above $\sim$10~GPa, both $T_{\rm N}$ and the valence value reach as high as the that of the isostructural YbNi$_{3}$Al$_{9}$ at ambient pressure. 

\begin{figure}[t]
\begin{center}
\includegraphics[width=0.5\textwidth]{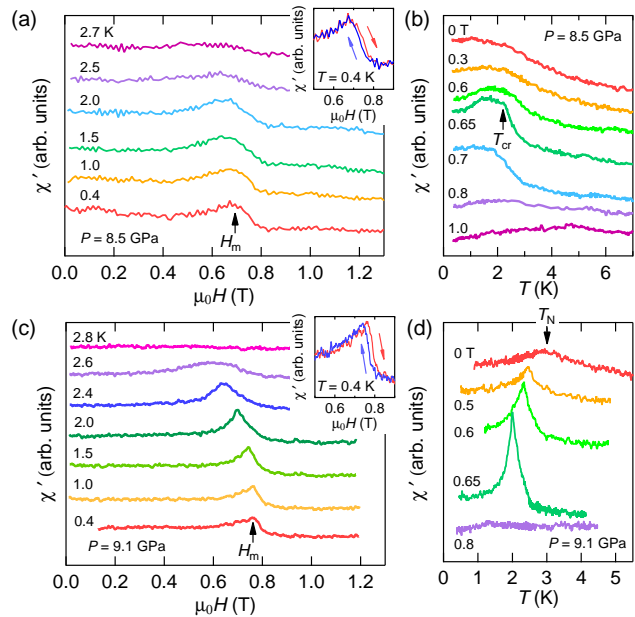}
\end{center}
\caption{(Color online) Magnetic field  and temperature dependence of $\chi_{\rm ac}^{'}$ in YbNi$_{3}$Ga$_{9}$ at 8.5 GPa (a), (b) and 9.5 GPa (c), (d). Curves are shifted for clarity. The upper insets in (a) and (c) are hysteresis curves of $\chi_{\rm ac}^{'}$($H$) measurements at $T$ = 0.4 K.%Arrows indicate the metamagnetic $T_{\rm m}$temperature and field $H_{\rm m}$.}
}
\label{Fig3}
\end{figure}

\begin{figure}[t]
\begin{center}
\includegraphics[width=0.5\textwidth]{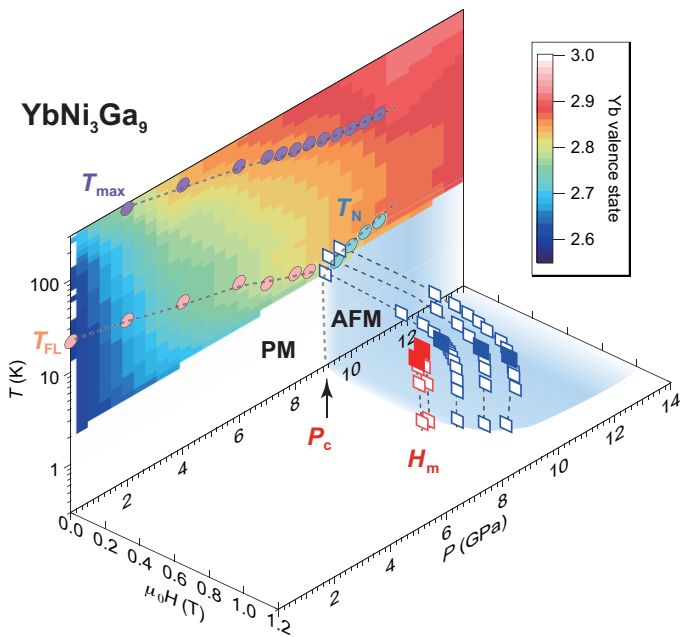}
\end{center}
\caption{(Color online) Contour plot of the Yb valence in the temperature-pressure phase diagram of YbNi$_{3}$Ga$_{9}$. The transition and crossover temperatures are deduced from resistivity (circles) and ac magnetic susceptibility (squares) measurements. 
Here, closed squares below and above $P_{\rm c}$ indicate the CP and TCP, respectively. The dashed lines are guides to the eye.}
\label{Fig4}
\end{figure}

	To search for the metamagnetic behavior, we focus on the effect of magnetic field in the vicinity of $P_{\rm c}$. The field and temperature dependences of $\chi_{\rm ac}^{'}$ for different constant temperatures and fields are displayed in Figs.~3(a),(b) and (c),(d) at below and above $P_{\rm c}$ ($\sim$9~GPa), respectively. At 8.5~GPa, by application of magnetic field along the $a$-axis, we found a first order metamagnetic transition in the $H$-sweep measurements at $H_{\rm m}$ $\sim$ 0.69 T with hysteresis at 0.4 K (see the inset of Fig.~3(a)). With increasing temperature, $H_{\rm m}$ slightly shifts to lower fields. As the temperature is increased further, the magnitude of the anomaly starts to decrease and is smeared out. Therefore, the first order metamagnetic transition becomes a crossover via a critical point (CP). As shown in Fig.~3 (b), the existence of the CP is also confirmed by the divergent behavior in the temperature dependence of $\chi_{\rm ac}^{'}$ at $T_{\rm cr}$ $\sim$ 2.1 K by tuning the magnetic field to $H$ $\sim$ $H_{\rm m}$ while $\chi_{\rm ac}^{'}$ exhibits a broad maximum away from $H_{\rm m}$. More interestingly, even above $P_{\rm c}$ at 9.1~GPa, a metamagnetic transition occurs from the AFM to a spin-polarized state at low temperatures (see Fig.~3(c) and (d)). The low temperature first order metamagnetic transition with a hysteretic behavior changes into the second order transition at higher temperatures through the tricritical point (TCP), where the distinct cusp in $\chi_{\rm ac}^{'}$ becomes sharper and enhanced in field and temperature dependence. This is reminiscent of the metamagnetic transition in YbNi$_{3}$Al$_{9}$ where the magnetic field is applied along the easy $a$-axis of magnetization \cite{sample_yamashita,Metamag_YbNi3Al9}, suggesting that the pressure-induced AFM phase in YbNi$_{3}$Ga$_{9}$ may have the helical magnetic structure identical to that of YbNi$_{3}$Al$_{9}$, i.e., the Yb magnetic moments are ferromagnetically aligned in the plane of Yb$_{2}$Al$_{3}$ layers, which has a weak interlayer magnetic coupling.

	The anomalies observed in the temperature and field scans in $\chi_{\rm ac}^{'}$ are summarized in the $H$-$T$-$P$ phase diagram of YbNi$_{3}$Ga$_{9}$ in Fig.~5 together with a contour plot of the Yb valence value in $T$-$P$ plane. Upon the application of pressure, we can see a clear evolution of the Yb-valence toward the magnetic trivalent state as well as the change from the nonmagnetic to the magnetic ground state. The striking feature of the phase diagram is that $T_{\rm FL}$ is cut off by a AFM transition temperature $T_{\rm N}$, suggesting the first-order nature of this transition at $P_{\rm c}$. Interestingly, the valence crossover region $\sim$2.8 converges toward $P_{\rm c}$. This characteristic variation of the Yb valence state in the $T$-$P$ phase diagram closely resembles that of YbInCu$_{4}$, which is known as a prototypical compound for the isostructural first-order valence transition (FOVT) between the high temperature phase with Yb$^{+2.97}$ and the low temperature phase with Yb$^{+2.84}$ at ambient pressure. In the case of YbInCu$_{4}$, a first-order ferromagnetic (FM) order emerges when the FOVT is suppressed to lower temperatures by applying pressure. Despite the difference of between FOVT and the valence crossover, these compounds as well as in other Yb-based heavy fermion compounds \cite{YbCu2Si2_valence,YbNiGe3_valence} share similar interplay between magnetic and valence instabilities. Hence, it is likely that a suppression of the magnetic order takes place due to enhanced valence fluctuations, giving rise to the occurrence of the first-order magnetic transition. In fact, the slave-boson mean-field calculation demonstrates that a coincidence of the AFM transition and the valence crossover at $T$ $\sim$ 0~K could occur depending upon the strength of the hybridization, causing the first-order AFM transition \cite{valence}.
	
	With approaching $P_{\rm c}$ from the paramagnetic side, there exists the first-order metamagnetic transition with the CP, which seems to merge to the TCP of the AFM order above $P_{\rm c}$. Here we consider the Clausius-Clapeyron relation for the metamagnetic field $H_{\rm m}$: $d$($\mu_{\rm 0}H_{\rm m}$)/$d$$T$~= ~-$\Delta$$S$/$\Delta$$M$, where $M$ and $S$ denote the magnetization and the entropy, respectively. Since the first order metamagnetic line below $P_{\rm c}$ has a negative slope in the $H$-$T$ plane, the entropy of the low-field region is smaller than that of high-field region. One possible interpretation of this result is that the Yb ion changes from the mixed-valent nonmagnetic state to the trivalent state with magnetic degrees of freedom by the application of magnetic field. Such a field-induced valence change indeed occurs in Yb compounds under high magnetic field \cite{YbInCu4_vacelnce_H}. Furthermore, recent theoretical calculations for an extended periodic Anderson model explain that the emergence of FOVT or the valence crossover is governed by the Coulomb repulsion between the $f$ and conduction electrons, and thus the FOVT is induced by applying the magnetic field even in the intermediate-valence state, resulting in the appearance of the metamagnetic behavior \cite{thery_valence_f}. It is worth noting that the enhanced ferromagnetic fluctuations is predicted to develop near the CP of FOVT \cite{theory-valence}, in accordance with our observation of the striking enhancement in $\chi_{\rm ac}^{'}$ near the CP of the metamagnetic line. Therefore we ascribe that the metamagnetic crossover at zero field evolves into a sharp first-order metamagnetic transition associated with the valence instability under magnetic field. We also speculate that the critical fluctuations due to the proximity to the valence crossover line extended from the FOVT line is a key ingredient responsible for the unconventional critical behavior, especially for the divergence of uniform susceptibility in paramagnetic phase, in other Yb systems such as YbRh$_{2}$Si$_{2}$ and $\beta$-YbAlB$_{4}$.
%Another possible explanation for smaller entropy below $H_{\rm m}$ is that the AFM short range ordering develops below $H_{\rm m}$, which is consistent with the emergence of the AFM ordering above $P_{\rm c}$. Considering the magnetic structure of YbNi$_{3}$Ga$_{9}$ inferred from that of YbNi$_{3}$Al$_{9}$, it is expected that application of magnetic field induces transfer from the AFM to FM correlations. 
For YbNi$_{3}$Ga$_{9}$, the extremely low value of the critical field $H_{\rm m}$ implies the closeness to a quantum critical endpoint (QCEP) of the FOVT, at which diverging valence fluctuations could be coupled to the Fermi-surface instability. It is an important experimental challenge to determine the location of the QCEP and the critical behavior by fine tuning pressure and magnetic field, which deserves further investigations.

	In conclusion, we present the phase diagram of YbNi$_{3}$Ga$_{9}$ as a function of pressure, magnetic field and temperature. We identify the clear Yb valence crossover toward $P_{\rm c}$ of the the pressure-induced AFM transition and, moreover, a first-order metamagnetic transition possibly due to the valence instability. The resulting phase diagram provides new insights into the unconventional quantum critical behavior in heavy fermion systems.

	We acknowledge discussions with K. Miyake, N.~K.~Sato and N.~Mori. We are grateful to H.~ Sumiya and T.~Irifune for providing the NPD anvils. This work is partially supported by Grants-in-Aid (No.24740220, 24540389) for Scientific Research from the Japan Society for the Promotion of Science, the approval of the Japan Synchrotron Radiation Research Institute (JASRI) (Proposal No.2011B2097, 2011B2094, 2011B2092, 2012A1283, 2012A1843,
2012B1976, 2012B0046, 2013A0046), and Grants-in-Aid for Scientific Research on Innovative Areas ‘‘Heavy Electrons’’ (No.20102007) from the Ministry of Education, Culture, Sports, Science and Technology, Japan.

\bibliography{ref.bib}

\end{document}